\documentstyle[twoside,fleqn,espcrc2]{article}

\def\Sym{{\cal S}}
\def\Si{\Sigma}
\def\Sib{\bar{\Sigma}}

\def\Qred{Q_{\!{}_{\rm red}}}
\def\Pred{P_{\!\!{}_{\rm red}}}
\def\Sred{S_{\!{}_{\rm red}}}
\def\H{{\cal H}}
\def\Obs{{\cal O}}

\def\Bbb{\bf}
\def\BbbR{{\Bbb R}}
\def\dotBbbR{\dot{\Bbb R}}
\def\BbbZ{{\Bbb Z}}
\def\rep{{\sf T}}
\def\sltwor{{{\rm SL}(2,\BbbR)}}
\def\sltwoz{{{\rm SL}(2,\BbbZ)}}
\def\psltwoz{{{\rm PSL}(2,\BbbZ)}}
\def\im{\hbox{Im}(h_F)}
\def\ker{\hbox{Ker}(h_F)}
\def\rpthree{{\BbbR}P^3}
\def\rptwo{{\BbbR}P^2} 
\def\semi{\mathbin{\hbox{\hskip2pt\vrule height 4.1pt depth -.3pt
                width .25pt \hskip-2pt$\times$}}}

\title{Determination and Reduction of Large Diffeomorphisms}

\author{D. Giulini\address{Fakult\"at f\"ur Physik der
           Universit\"at Freiburg,\\
           Hermann-Herder Stra\ss e 3, D-79104 Freiburg, Germany}}%

\begin{document}

\begin{abstract}
For diffeomorphism invariant theories we consider the problem of
how to determine and reduce diffeomorphisms which are not
in the identity component.
\end{abstract}

\maketitle

\section{THE PROBLEM}

We shall consider diffeomorphism invariant theories within the 
Hamiltonian formulation, where space-time is assumed to be a 
topological product $M\cong\Si\times\BbbR$. The constraints of
the theory will only generate the identity component
$D^0(\Si)$ of some subgroup $D(\Si)$ of diffeomorphisms of $\Si$.
This means that after a reduction by $D^0(\Si)$, i.e.,
implementing the constraints -- which itself is a highly
non-trivial problem --, one still has a residual action of the
discrete and generally non-abelian and infinite group
\begin{equation}
\Sym(\Si):=D(\Si)/D^0(\Si)
\end{equation}
on the reduced state space $\Sred$. In what follows, we shall
always restrict to orientable $\Si$ and orientation
preserving $D(\Si)$. But this is not essential.
In principle one is now free to regard $\Sym(\Si)$ either as
residual part of the gauge group, i.e., as redundancy, or as
proper physical symmetry. In the first case physical observables
must lie in the commutant of the von Neumann algebra generated by
$\Sym(\Si)$, whereas in the second {\it some} physical observables
must break the symmetry to render it observable. In the
first case the reduction procedure is not completed and we must
consider the set of $\Sym(\Si)$-orbits in $\Sred$ as faithful space of
physically distinguishable states. Whether this orbit space can be
given a sufficiently well behaved structure will depend crucially
on the details on $\Sym(\Si)$'s action on $\Sred$. 
A priori there is absolutely
no reason to expect the structure and action of a general infinite
discrete group to be nice. Also, the analogous problem to the one
posed by continuous spectra of the generators of continuous Lie
groups also occurs even for the simplest infinite discrete groups.
This will be seen in the first example.
In addition, iff a discrete group is not an abelian extension of
a finite group it is not of type I, meaning that general
representations may be written in different, mutually disjoint
direct integral decompositions of irreducibles. Whether this
really implies difficulties in the quantum theory, for example
as ambiguities in the determination of sectorial structures
by the reduction process, is presently not known to us.

\section{FIRST EXAMPLE}

In this section we wish to illustrate some typical problems
connected with the {\it reduction} of large diffeomorphisms.

Einstein gravity in 2+1 dimensions can be considered as $ISO(2,1)$ 
Chern-Simons gauge theory \cite{Witten}. The point of doing this
is that the constraints can be solved and the reduced state space 
be constructed. We specialize to $\Si=T^2$, where $T^2$ denotes
the two-torus. In the so-called metric sector the classical reduced 
state- (or phase-) space $\Pred$ is a cotangent bundle $T^*(\Qred)$, 
where the reduced configuration space $\Qred=\dotBbbR^2 / \BbbZ_2$ 
is the punctured plane 
($\dotBbbR^2=\BbbR^2-\{0\}$) with antipodal points identified by 
the ${\BbbZ}_2$-action of reflections at the origin. Let 
$\vec q=(q_1,q_2)$ be cartesian coordinates on $\dotBbbR^2$ and 
$\vec p=(p_1,p_2)$ the conjugate momenta. Their interpretation is as 
follows: Let $\alpha$ and $\beta$ be two closed curves on $T^2$ 
whose homotopy classes generate $\pi_1(T^2)\cong\BbbZ\times \BbbZ$. 
Then $(q_1,q_2)$ and $(p_2,-p_1)$ are $ISO(2,1)$ holonomies along 
$(\alpha, \beta)$, which are boosts in $x$-direction for the $q$'s 
and spatial translations in $y$-direction for the $p$'s. 
See e.g. \cite{Don-Jorma} for details. If $D(T^2)$ denotes the
(orientation preserving) diffeomorphisms, one
has $\Sym(T^2)=D(T^2)/D^0(T^2)\cong \sltwoz$. Its action
on $\Qred$ is just the projection of the defining representation
on $\BbbR^2$, so that on $\Qred$ only
$\sltwoz/{\pm 1}=\psltwoz\cong Z_2*Z_3$ ($*=$ free product)
acts effectively. On $\Pred$ the action is just given by canonically
lifting the action on $\Qred$. The action of $\sltwoz$ on $\dotBbbR^2$
is wild indeed. For example, the stabilizer subgroups of points $(x,y)$
with irrational $y/x$ are trivial whereas they are $\cong {\bf Z}$
for rational $y/x$. Since points with rational slopes lie dense,
the isomorphicity classes of stabilizer subgroups are nowhere locally
constant and hence the quotient is nowhere even locally a manifold.
However, the lifted action on $T^*(\dotBbbR^2)$ is free and
properly discontinuous on the open and dense set
$\{(\vec q,\vec p)\,\vert\, \vec q\cdot\vec p\not =0\}$

Due to the fact that our classical phase space is a cotangent
bundle, the quantization of this model may naturally proceed
with the realization of the Hilbert space as $L^2$-functions 
on $\Qred$:
\begin{equation}
\H=L^2_+(\dotBbbR^2; dq_1dq_2),
\end{equation}
where $+$ indicates that the functions must be invariant under
$\vec q\rightarrow -\vec q$. The choice of the Lebesgue measure
is not arbitrary: $ISO(2,1)$ is the cotangent-bundle-group of $SO(2,1)$
and the space of flat $ISO(2,1)$ connections is the cotangent bundle
over the space of flat $SO(2,1)$ connections. The latter has itself
a natural symplectic structure whose associated Liouville measure
one may generally use in 2+1 dimensions to define the inner product
(see p. 272 and 276 of \cite{AAbook}). Applied to our case this results
in $dq_1dq_2$. Its $\sltwoz$-invariance implies that we have a unitary
representation $\Sym(T^2)\times\H\rightarrow \H$ given by 
$\rep:\ ([g],\psi)\mapsto \rep_{[g]}\psi:=\psi\circ g^{-1}$,
where $g\in\sltwoz$ is any preimage of $[g]\in\psltwoz$ under the
natural projection. It is possible to explicitly decompose this
representation into a direct integral of
unitary irreducibles \cite{Jorma-Nico}:
\begin{eqnarray}
\rep &= \int_{-\infty}^{\infty}ds\, \rep_s  \\
  \H &= \int_{-\infty}^{\infty}ds\, \H_s,   
\end{eqnarray}
where $\H_s\cong L^2(S^1,d\phi)$. The irreducibles $\rep_s$ are
restrictions to $\sltwoz$ of the irreducibles $C^0_q$ of $\sltwor$
from the principal series with Casimir invariant $q=(s^2+1)/4$.
The interesting features of this decomposition are
1.) the absence of the trivial representation,  2.) the absence 
of finite dimensional irreducibles, 3.) the purely continuous 
Casimir spectrum. There are no irreducible subspaces in $\H$ but
closed invariant subspaces are given by 
$\H_{\Delta}=\int_{\Delta}ds\,\H_s$ for any measurable set
$\Delta\subset\BbbR$.

This situation is well known from continuous groups with
continous spectra of their generators. Consider for example 
the translations in y-direction acting on $L^2(\BbbR^2,dxdy)$.
But in this example the trivial representation occurs in the direct
integral decomposition. So if we wanted to interpret the y-translations
as gauge redundancy we could identify the reduced quantum state space
with the integrand $\cong L^2(\BbbR,dx$) carrying the trivial
representation. This is just what the procedure of ``group
averaging'' leads to \cite{AAetal}. In our example such a simple 
identification does not seem possible. Note that if we
just kept $\H$ as state space and implemented the unobservability
of the transformations in $\Sym$ by restricting the algebra of
observables $\Obs$ to the commutant $\{\Sym(T^2)\}'$ of $\Sym(T^2)$
in $B(\H)$ (bounded Operators), then $\H$ would not contain a single
pure state for $\Obs$. The proof is simple: Let $\psi\in\H_{\Delta}$
be normalized. We can always find disjoint measurable sets 
$\Delta_{1,2}$ of non-zero measure such that 
$\Delta=\Delta_1\cup\Delta_2$, hence
$\H=\H_{\Delta_1}\oplus\H_{\Delta_2}$, and associated decomposition
$\psi=\lambda_1\psi_1+\lambda_2\psi_2$ with normalized $\psi_{1,2}$.
Since each $\H_{\Delta_i}$ reduces $\Obs$, the density matrices
$\rho=\vert\psi\rangle\langle\psi\vert$ and
$\rho_{1,2}=\vert\psi_{1,2}\rangle\langle\psi_{1,2}\vert$ obey
$\rho=\vert\lambda_1\vert^2\,\rho_1+\vert\lambda_2\vert^2\,\rho_2$
as linear functionals on $\Obs$. Hence $\rho$ is a non-trivial
convex sum and therefore a mixed state on $\Obs$.

We ask: 1.) regarding $\Sym(T^2)$ as a proper physically 
symmetry group implies observables outside $\{\Sym(T^2)\}'$. 
How could these be justified physically?
2.) Regarding $\Sym(T^2)$ as part of the gauge group
necessitates further reduction. How?

\section{THE GENERAL 3+1 CASE}

Except in cosmology we are usually not interested in closed
3-manifolds representing space. Appropriate for the description of
isolated gravitating configurations are 3-manifolds $\Si$ with one
regular end, i.e., there exists a compact set $K\subset\Si$ so that
$\Si-K$ is homeomorphic to $\BbbR\times S^2$. This is precisely the
condition that the one-point-compactification $\Sib=\Si\cup\infty$ ($\infty$
is the added point) is again a manifold. The mapping class group
we are interested in is then conveniently characterized by using
the fiducial manifold $\Sib$ (e.g. \cite{Nico2}). Let $D_F(\Sib)$
be the diffeomorphisms of $\Sib$ that fix the frames at $\infty$
and $D_F^0(\Sib)$ its identity component. We define
\begin{equation}
\Sym(\Si):=D_F(\Sib)/D_F^0(\Sib).
\end{equation}

One studies $\Sym(\Si)$ by considering the group homomorphism
\begin{eqnarray}
h_F:\Sym(\Sib)\rightarrow \hbox{Aut}(\pi_1(\Sib,\infty))\\
h_F([\phi])([\gamma]):=[\phi\circ\gamma],
\end{eqnarray}
where $\gamma$ is a loop based at $\infty$, $[\gamma]$ its homotopy
class, $\phi\in D_F(\Sib)$, and $[\phi]$ its class in $\Sym(\Si)$.
The strategy is to obtain $\Sym(\Si)$ from 1.)~$\ker$ = kernel of
$h_F$, 2.)~$\im$ = image of $h_F$, 3.)~a prescription
to extend $\im$ by $\ker$. Given the connected sum decomposition
of $\Sib$, it is indeed possible to explicitly present $\Sym(\Sib)$
for a large class of 3-manifolds. The generating diffeomorphisms fall
into three classes: 1.) internal-, 2.) exchange -, and 3.) slide
diffeomorphisms. To explain this we recall that any
compact orientable 3-manifold is uniquely built as finite connected
sum of so-called prime manifolds (see \cite{Nico1} for details and
references): $\Sib=P_1\uplus\cdots\uplus P_n$. Then
$\pi_1(\Sib)\cong \pi_1(P_1)*\cdots *\pi_1(P_n)$.
In this way $\Si$ is represented by a 3-disk $B$ to which primes
$P_i\not = S^1\times S^2$ are glued by removing an open 3-disk
from $P_i$ and $B$ and identifying the resulting 2-sphere boundaries
so as to match the given orientations. If $P_i=S^1\times S^2$ we
remove two open 3-disks from $B$ and identify the boundaries left
with the boundary 2-spheres of $[0,1]\times S^2$. We thus view $\Si$
as a configuration of $n$ elementary objects connected to the base
$B$ by 2-spheres, like particles with internal structure ``moving''
in $B$. Internal diffeomorphisms are those which (up to isotopy)
have support within the primes. Exchanges are those non-internal ones
that leave $B$ and the interiors of the $P_i$'s setwise invariant
(i.e. permutations of diffeomorphic primes). Finally, slide
generators account for the fact that primes can penetrate and
``move'' through each other. They mix exterior and interior points.
Roughly speaking, each connecting sphere of a prime can be slid a
full turn within a closed tube whose axis-loop generates an element
of the fundamental group of another prime. Homotopic loops define
isotopic slides. Slides form an invariant subgroup $G^S\subset\Sym(\Si)$
(see \cite{Nico1} and its references).

Now, the so-called Fuks-Rabinovich presentation for
$\hbox{Aut}(G_i*\cdots *G_n)$ allows to explicitly present $\im$,
once we have presentations for for each $\Sym(P_i)$
(see \cite{Banach} and its references). Given this, we obtain
a presentation for $\Sym(\Si)/\ker$. The problem is now to
determine $\ker$ and the way it extends $\im$. If all $P_i$
satisfy that homotopic diffeomorphisms are also isotopic
(no prime violating this seems to be known) and no $P_i$ is a
homotopy sphere then it is known that $\ker$ consists of
rotations parallel to connecting spheres \cite{McCullough}.
Then $\Sym(\Si)=\ker\semi\im$ where only the permutations in
$\im$ act non-trivially (in the obvious way) on $\ker$.
In this way presentations for connected sums of an arbitrary
number of $\rpthree$'s or an arbitrary number of $S^1\times S^2$'s
were obtained \cite{Banach} in terms of three and four 
generators respectively.

The obvious semi-direct product of the internal symmetry
group $G^I=\Sym(P_1)\times\cdots\times\Sym(P_n)$ and the
permutations form the so-called ``particle subgroup''
$G^P\subset \Sym(\Si)$ \cite{Banach}. $G^P$ and $G^S$ together
exhaust $\Sym(\Si)$ but they may intersect non-trivially.
It has been shown that iff $P_i\not =S^1\times S^2\,\forall i$,
$G^P\cap G^S=\{e\}$ and $\Sym(\Si)\cong G^S\semi G^P$, and 
that $G^S$ is perfect if more than two primes are $S^1\times S^2$
\cite{Banach}.

\section{SECOND EXAMPLE}

We consider the connected sum of two real projective spaces 
$\Sib=\rpthree\uplus\rpthree$ to illustrate the determination of
$\Sym(\Si)$ according to the general scheme outlined above.

One way to understand the manifold $\Sib$ is to look at the fundamental 
domain $F=\{\vec x\in\BbbR^3\mid 1\leq\Vert\vec x\Vert\leq3\}$.
We label points in $F$ by $(r,\vec n)$ where $r=\Vert\vec x\Vert$
and $\vec n=\vec x/r$. Let $S_{r'}$ denote the sphere $r=r'$ 
and $\sigma({\vec n}')\subset F$ the radial segment $\vec n={\vec n}'$.
To obtain $\Sib$ we identify antipodal points on $S_1$ and on $S_3$.
The sets $C(\pm\vec n):=\sigma(\vec n)\cup\sigma(-\vec n)$ define 
a $\rptwo$ worth of circles which establishes $\Sib$ as circle 
bundle over $\rptwo$ (which is not principal). $S_2$ may be taken as
the sphere along which the connected sum of the two $\rpthree$'s is taken.
Obviously it cuts each fiber twice. We also place $\infty$ on $S_2$, 
say at $\vec n=2{\vec e}_z$.

We have 
$\pi_1(\Sib)=\BbbZ_2*\BbbZ_2=\{a,b\mid a^2=e=b^2\}$, where $e$ denotes 
the identity. $a$ and $b$ may be generated by meridians on $S_1$ and 
$S_3$ respectively. One then sees from a picture of $F$ that a circle
fibre generates $ab$, which itself generates a subgroup
$\cong\BbbZ\subset\pi_1(\Sib)$. The conjugacy class of $ab$ in
$\pi_1(\Sib)$ consists
only of $ab$ and $ba=(ab)^{-1}$. The map $ab\rightarrow ba$ generates 
$\hbox{Aut}(\BbbZ)\cong \BbbZ_2$ and may be interpreted as the action 
of the fundamental group of the base $\rptwo$ on the fundamental group 
of the fibre $S^1$. In fact, $\pi_1(\Sib)$ is the semi-direct 
product of these two groups: $\BbbZ_2*\BbbZ_2\cong \BbbZ\semi\BbbZ_2$.
To give an explicit isomorphism let $(n,p)\in \BbbZ\semi\BbbZ_2$
with $p\in \{1,-1\}$ (multiplicative notation).
Then $(n',p')(n,p)=(n'+p'n,p'p)$ and an isomorphism 
$\phi: \BbbZ\semi\BbbZ_2\rightarrow \BbbZ_2*\BbbZ_2$ may be defined 
by $\phi(n,1)=(ab)^n$ and $\phi(n,-1)=(ab)^na$. One easily verifies the 
homomorphism property. In- and surjectivity are obvious.

It is known that $\Sym(\rpthree)=\{e\}$ so that $G^P\cong\BbbZ_2$ 
is generated by exchanging the two primes. In $F$ the exchange can be 
defined by a reflection at $S_2$ followed by a reflection at the 
yz-plane. This still rotates tangent vectors at $\infty$ by $\pi$ 
in the $y$ direction, but a slight modification by rotating back a 
small 3-disk about $\infty$ renders this diffeomorphism an element of
$D_F(\Sib)$. It defines a generator $\omega$ of $\Sym(\Sib)$ satisfying 
$\omega^2=e$. There are two slides, $\mu_{12}$ and $\mu_{21}$, 
corresponding to sliding the second through the first prime along 
some generator of its fundamental group and vice versa.
It sufficies to define one of them: In $\Sib$ consider the
closed solid tori
$T_{1,2}=F\cap\{\vec x\in \BbbR^3\mid y^2+z^2\leq R_{1,2},
\, 1<R_1<R_2<2\}$ and a diffeomorphism with support in
the closure of $T_2-T_1$ that slides $T_1$ against $T_2$
a full turn parallel to their common axis. This defines a slide
of the ``inner'' ($r<2$) through the ``outer'' ($r>2$) prime.
Since the prime's fundamental group is $\BbbZ_2$, we have
$\mu_{12}^2=e=\mu_{21}^2$.
The Fuks-Rabinovich presentation implies that there 
is no other relation between the slides (this would change if we 
considered more than two primes \cite{Banach}). Hence 
$G^S\cong \BbbZ_2*\BbbZ_2$.  Finally $\Sym(\Sib)=G^S\semi G^P$ where 
$G^P$'s action on $G^S$ is $\omega\mu_{12}\omega^{-1}=\mu_{21}$.
We can use this last relation to eliminate $\mu_{21}$ from the 
presentation and just retain $\omega$ and $\mu=\mu_{12}$ with no 
other relation except their idempotency. Hence 
\begin{equation}
\Sym(\Sib)\cong \BbbZ_2*\BbbZ_2\cong \BbbZ\semi\BbbZ_2
\end{equation}
As shown above, the generators of $\BbbZ$ and $\BbbZ_2$ in the 
semi-direct product may be identified with $\omega\mu$ and $\mu$  
respectively. Compared to the group $\BbbZ_2*\BbbZ_3$ considered 
in the first example, $\BbbZ_2*\BbbZ_2$ has a much simpler 
representation theory being a semi-direct product.
(It is clearly of type I by the criterion mentioned above).
There are the obvious four one-dimensional irreducible
representations and the more interesting one-parameter
($0<t<\pi$) family of two-dimensional ones given by
$\omega\mapsto\tau_3$ and $\mu\mapsto\tau_1\sin t+\tau_3\cos t$,
where $\tau_i$ are the Pauli Matrices \cite{Aneziris}.

\end{document}